\documentclass[twocolumn,showpacs,showkeys,preprintnumbers,amsmath,amssymb]{revtex4}

\usepackage{graphicx}
\usepackage{dcolumn}
\usepackage{bm}
 
\begin{document}

\title{Sub-Kelvin Cooling of a Macroscopic Oscillator and femto-Newton Force Measurement}

\author{F. Mueller}
\author{S. Heugel}
\author{L. J. Wang} 
\email{lwan@optik.uni-erlangen.de}
\affiliation{Max Planck Research Group, Institute of Optics, Information and 
Photonics, University Erlangen-Nuremberg, D-91058 Erlangen, Germany}
\homepage{http://www.optik.uni-erlangen.de}

\date{September 28, 2007}

\begin{abstract}
Measuring very small forces~\cite{brag1_1977,chen1_1993,roll1964,brag2_1972,baessler1999,abramo1992,gillies1992,chen2_1990,saulson_1990,ritter1_1999,ritter2_1985,luther1982,quinn2001,schwarz1998,niebauer1995}, particularly those of a gravitational nature, has always been of great interest, as fundamental tests of our understanding of the physical laws. 
Ultra-long period mechanical oscillators, typically used in such measurements, will have $k_{B}T/2$ of thermal energy associated with each degree of freedom, owing to the equal-partition of energy. 
Moreover, additional seismic fluctuations in the low frequency band can raise this equivalent temperature significantly to $\sim10^{5}\,K$. 
Recently, various methods using opto-mechanical forces~\cite{cohadon1999,metzger2004,gigan2006,arcizet1_2006,kleckner2006,arcizet2_2006} have been reported to decrease this thermal energy for $MHz$, micro-cantilever oscillators, effectively cooling them. 
Here we show the direct, dynamical cooling of a gram-size, macroscopic oscillator to $\sim300\,mK$ in equivalent temperature - noise reduction by a factor of $\sim10^{6}$. 
By precisely measuring the torsional oscillator's position, we dynamically provide an external 'viscous' damping force. 
Such an added, dissipative force is essentially free of noise, resulting in rapid cooling of the oscillator. 
Additionally, we observe the time-dependent cooling process, at various cooling force parameters. 
This parameter dependence agrees well with a simple physical model which we provide. 
We further show that the device is sensitive to forces as small as $<100\,fN$ - a force only a few percent of that typically exerted by a single biological molecule or that observed in a typical gravity experiment. 
We also demonstrate the dynamic control of the oscillator's natural frequency, over a span of nearly two decades. 
The method may find important applications in precision measurements of very weak forces.
\end{abstract}


\maketitle
Ultra-long period mechanical oscillators have important applications in precision measurement science, both for measuring very small forces and for providing a non-moving reference point in interferometry. 
The measurement of very small forces~\cite{brag1_1977,chen1_1993,roll1964,brag2_1972,baessler1999,abramo1992,gillies1992,chen2_1990,saulson_1990,ritter1_1999,ritter2_1985,luther1982,quinn2001,schwarz1998,niebauer1995}, particularly those of a gravitational nature, has always been of great interest, for these measurements often provide tests of our fundamental understanding of the physical laws, such as the \lq weak\rq~equivalence principle~\cite{roll1964,brag2_1972,baessler1999} and gravitational waves~\cite{abramo1992}. 
The experiments usually employ low-frequency, macroscopic, mechanical oscillators, such as a torsion balance~\cite{gillies1992} or a suspended mass~\cite{abramo1992}. 
As a result of the very small restoring force $k$ and a relatively large mass $m$ of the system, high sensitivity often implies low natural frequency. However, the micro-seismic noise, and more fundamentally, thermal noise~\cite{chen2_1990,saulson_1990,ritter1_1999,ritter2_1985} pose important limits on the sensitivity of these devices.
 
In another group of experiments involving precision measurements of distance or Earth gravity and more generally in interferometry, a common limitation is the uncertainty of the position of the \lq reference point\rq~. 
Since distance is a relative quantity from a reference point, fluctuations of the reference point during the measurement time will pose a limit on precision. 
However, to obtain an absolutely non-moving, stationary point is no easy task - the only practical solution being to use a very long period oscillator such that during a short data acquisition time, the oscillator is in effect stationary. 
For example, in free-fall, absolute gravimeters~\cite{schwarz1998,niebauer1995} where Earth gravity $g$ can be measured to the $10^{-9}$ accuracy, a long-period oscillator is often employed as the reference point of the interferometer. 
Almost universally, external noise in these devices is an important issue. 
Thermal noise, being one that is particularly difficult to overcome, is a result of coupling with the \lq noisy\rq~external environment via dissipation, causing the system to fluctuate with a thermal energy of $k_{B}T/2$. 
In this letter, we show how this fluctuation can be reduced via a direct, dynamic \lq cooling\rq~method. 
We further show how the oscillator's natural frequency can be dynamically controlled, and demonstrate such control over a range of almost two decades. Finally, we note that cooling has been applied to micro-cantilever devices (with $\sim MHz$ natural frequency) in recent years, often using opto-mechanical forces~\cite{cohadon1999,metzger2004,gigan2006,arcizet1_2006,kleckner2006,arcizet2_2006}. 
Much of this group of work aimed at reaching the quantum limit, or even entanglement of these devices~\cite{gigan2006,kleckner2006} with a laser beam. We note here that the quantum limit is frequency dependent, and is entirely negligible in the present work, using a gram-size mass. 
Unlike the micro-cantilevers with relatively stiff arms, a torsion oscillator is typically more responsive to small forces.

\begin{figure}                                          
\includegraphics{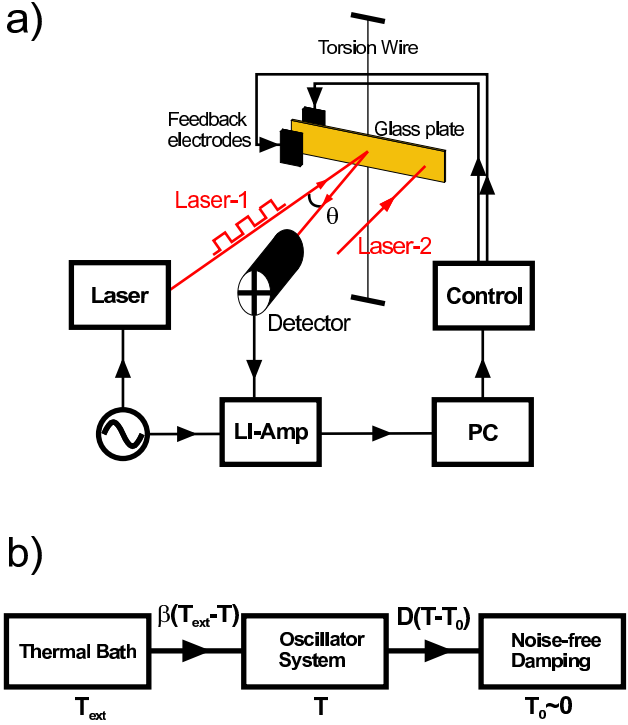}
\caption{Schematic experimental setup. The oscillator's angular position is measured by pointing Laser-1 to the center of the oscillator and detecting the reflected light beam with a low-noise, four-quadrant position detector. Laser-2 is used to exert a known radiation pressure force on one arm of the balance, causing an angular deflection. Two capacitive feedback electrodes on the left side are used for dynamic control. High sensitivity lock-in detection of the balance's angular position~\cite{lorrain1991} and dynamic control loop. LIA is the Lock-in amplifier. (b) Flow of thermal fluctuation energy.}
\label{fig1}
\end{figure}

We begin by considering a simple, mono-mode torsion oscillator, as shown in Fig.~\ref{fig1} a.
The equation of motion is
\begin{equation}
\label{equ1}
\ddot{\theta}+\beta\dot{\theta}+\omega_{0}^{2}\theta=\alpha_{th}(t)+\alpha_{c}(t),
\end{equation}
where $\theta$ represents the oscillator's angular position, $\omega_{0}=\sqrt{\tau/I}$ the natural frequency, and $\beta$ the damping rate due to coupling with the environment. 
The oscillator is under the influences of two external angular accelerations.
The acceleration $\alpha_{th}(t)$ due to a thermal force fluctuates randomly and can be assumed to have little correlation at the time scale of the oscillation period, with essentially a white spectrum: $\left\langle \alpha_{th}(t)\alpha_{th}(t')\right\rangle=K\delta(t-t')$, where $K$ is a constant. 
At the time scale of $\beta^{-1}$, the oscillator reaches thermal equilibrium with the thermal environment at temperature $T_{ext}$.

The external control acceleration, $\alpha_{c}(t)$ can take a more general form that is a function of time, as well as the angular position and its velocity or acceleration. 
Furthermore, this control acceleration can also be a nonlinear function, thus enabling almost arbitrary control of the oscillator. 
Here, to demonstrate cooling, we employ a simple viscous damping: $\alpha_{c}(t)=-D\dot{\theta}$, where $D$ is the damping rate. 
Applying this external control acceleration, it is easy to see how the system will evolve.
 
By multiplying both sides of Eq.~\ref{equ1} by $\dot{\theta}$ and taking an ensemble average, and noting that $\left\langle \dot{\theta}\right\rangle=\omega_{0}^{2}\left\langle \theta^{2}\right\rangle=k_{B}T/I$, we have
\begin{equation}
\label{equ2}
\frac{dT}{dt}+\beta T=-DT+\frac{I}{k_{B}}\left\langle \dot{\theta}\cdot \alpha_{th}(t)\right\rangle.
\end{equation}
Here, $T$ is the system's temperature at time $t$. The last term in Eq.~\ref{equ2}, $\left\langle \dot{\theta}\cdot \alpha_{th}(t)\right\rangle$, has an explicit physical meaning of heating by energy exchange with the environment. 
It is easy to show that $\left\langle \dot{\theta}\cdot \alpha_{th}(t)\right\rangle=\beta k_{B}T_{ext}/I$.
At thermal equilibrium, when the system temperature $T$ is stationary, we have
\begin{equation}
\label{equ3}
T=\frac{\beta}{\beta+D}T_{ext}.
\end{equation} 
Eqs.~\ref{equ2} and~\ref{equ3} afford a very clear physical picture - being in contact with both the external thermal environment and the noiseless (zero-temperature) cooling force, the system's overall rates of heating $\beta\cdot(T_{ext}-T)$ and cooling $D\cdot T$ cancel each other at the equilibrium (Fig.~\ref{fig1} b).

In the experiment, Fig.~\ref{fig1} a, we use a macroscopic torsional oscillator made of a glass substrate $50\,mm\times10\,mm\times0.15\,mm$ in size 
($\sim0.2\,g$). 
The substrate is gold-coated on both sides and fixed to a $15\,cm$ long, $25\,\mu m$ thick tungsten wire. 
The oscillator has a moment of inertia $I=4.4\times10^{-8}\,kg\times m^{2}$ and a torsional constant $\tau=2.2\times10^{-7}\,Nm\,rad^{-1}$.
The pendulum oscillates at a natural frequency $f_{0}=0.36\,Hz$  with a quality factor of $Q\approx2,300$.
Using a laser beam reflected from the center of the oscillator and a high-sensitivity quadrant diode detector followed by a lock-in detection technique, we measure the oscillator's angular position with an accuracy of $\sim2\,nrad\,Hz^{-1/2}$. 
Reflecting from the center, the laser beam used in the optical lever exerts a negligible net torque. 
Two polished electrodes are positioned at one end of the balance. 
By electrically grounding the oscillator and changing the voltages on the electrodes, an electrostatic force can be applied to the oscillator in both directions. 
At the other end of the oscillator, a laser beam of controllable intensity can be reflected from the gold-coated surface, to introduce a well-calibrated light force. 
The entire apparatus is kept in a high vacuum ($10^{-7}\,mbar$) environment which is mounted on top of an active vibration isolation system. 
In addition, seismic noise above the natural frequency is filtered from the oscillator by a special damping suspension.

The experiment relies on real-time, high accuracy determination of the oscillator's instantaneous angular position~\cite{lorrain1991}. 
To achieve this, we implement the detection system shown in Fig.~\ref{fig1} a. 
A laser beam is modulated at $100\,kHz$, and the reflected beam is detected by a low noise quadrant position detector. 
The resulting differential photoelectric current is first pre-amplified and subsequently demodulated using a lock-in amplifier. 
Using this technique, a tremendous sensitivity of $<20\,nrad/V$ is reached. 
This angular voltage signal is then digitized at a sampling rate of $5\,kHz$ and used as the input of a computerized digital control loop. 
Both input and output channels show analog/digital conversion noise levels of only $\sim20\,V\,Hz^{-1/2}$. 
Owing to this conversion noise, the closed-loop control system shows an equivalent noise level of $\sim2\,nrad\,Hz^{-1/2}$. 
Furthermore, the system is calibrated and tested to ensure that the \lq in-loop\rq~and \lq out-of-loop\rq~performance are in good agreement.

The digital loop has the advantage of convenient adjustment of control parameters. 
In the experiment, only proportional ($P$) and differential ($D$) controllers are used.
\begin{figure}                                          
\includegraphics{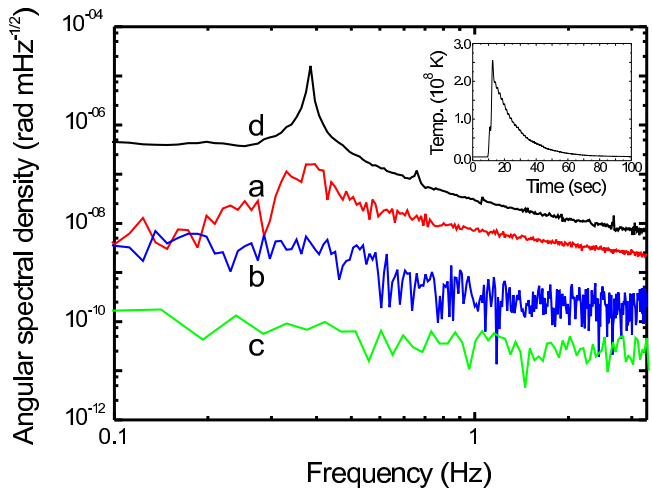}
\caption{The oscillator's position is cooled in three consecutive steps, shown in the measured power spectral density functions (a-c). For low damping (a), the equivalent noise temperature is $\sim6.8\times10^{4}\,K$, due to micro-seismic excitation. For a higher damping rate (b), this temperature is reduced to $\sim8\,K$, and finally in (c) to $\sim300\,mK$, which corresponds to a $4.3\,nrad$ RMS angular uncertainty. (d) An artificial electronic impulse applied to the feedback electrodes excites the oscillator to an equivalent temperature of $3\times10^{8}\,K$. In this case, the damping force is first disabled. After a few oscillation cycles, a low damping is applied. This causes an exponential decrease of the system's temperature, as shown in the inset.}
\label{fig2}
\end{figure}
Fig.\ref{fig2} shows the power density of the oscillator's spectra at various equivalent temperatures. 
First, the oscillator is freely running and is well coupled to the surrounding seismic noise near its natural frequency of $0.36\,Hz$ (curve a). 
Owing to this excess noise, the equivalent noise temperature is $\sim10^{5}\,K$, more than two orders of magnitude higher than the thermal noise at room temperature. 
Applying a dynamic damping control, the equivalent temperature can be cooled to various levels (curves b, c), depending on the setting of this damping parameter. 
The lowest temperature reached in the experiment is $<300\,mK$. 

To further examine the cooling process, we artificially excite the oscillator with an impulse applied to the control electrodes. 
After a few cycles, we measure the noise power spectrum of this excited oscillator (curve d) and it gives an equivalent temperature of $\sim3\times10^{8}\,K$. 
Then a low damping is applied, causing cooling. 
The inset of Fig.\ref{fig2} shows the time dependence of this cooling process, in good agreement with the prediction of the simple model given in Eq.~\ref{equ2}.

\begin{figure}                                          
\includegraphics{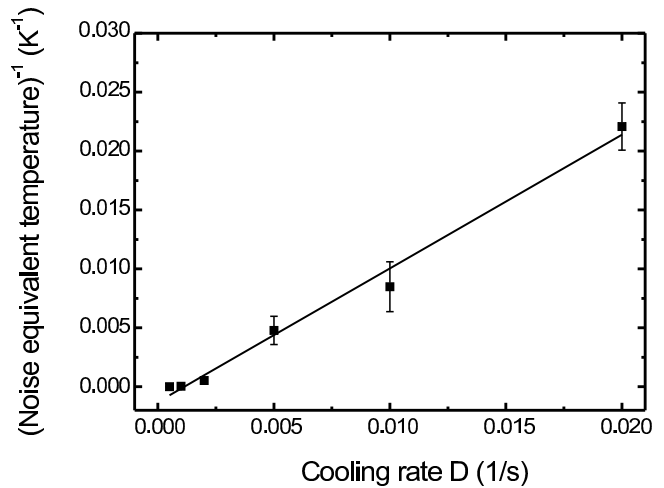}
\caption{Measured inverse equilibrium temperature of the oscillator versus varied damping rates. The result is in good agreement with Eq.~\ref{equ3}.}
\label{fig3}
\end{figure}

Eq.~\ref{equ3} gives a simple relation of the oscillator's equilibrium temperature when a certain cooling parameter $D$ is applied, based on our theoretical model. 
To further verify this model, we measure the oscillator's power spectrum and the equivalent temperature $T=I\omega_{0}^{2}\left\langle \theta^{2}\right\rangle/k_{B}$. 
Fig.\ref{fig3} shows the dependence of the equivalent temperature on the cooling rate; the theoretical and experimental results are in good agreement.
\begin{figure}                                          
\includegraphics{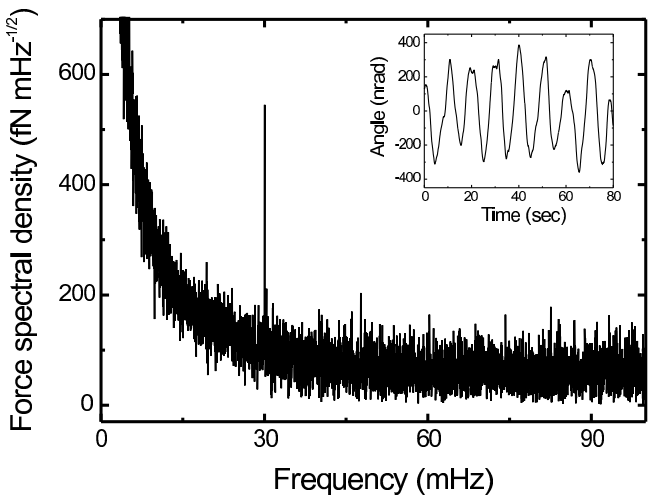}
\caption{A \lq cooled\rq~oscillator is sensitive to very weak forces. Here we apply a light force caused by a laser beam of only $50\,\mu W$ in amplitude ($30\,mHz$); this influence is clearly visible over a background of $\sim100\,fN\,mHz^{1/2}$. Inset shows the cooled oscillator's temporal response to the light force ($\pm15\,pN$) caused by laser light of $\sim5\,mW$ of peak-to-peak modulation.}
\label{fig4}
\end{figure}
Cooling or reducing noise of the oscillator can improve the sensitivity of measurement of very small forces. 
Although as in almost all cases the signal-to-noise ratio is not increased, by reducing noise, we can greatly reduce the dynamic range of the angular position detector and hence increase its sensitivity and gain. 
As an example, we show the detection of a very small force at the $fN$ level, caused by the radiation pressure of a $\mu W$ laser beam. 
To do this, a laser beam (laser-2 in Fig.~\ref{fig1} a) is directed onto and reflected from the gold-coated surface of the oscillator. 
The laser beam is modulated in a sinusoidal form at a period of $\sim33\,s$ with amplitude of $50\,\mu W$. 
This corresponds to a force of $330\,fN$ in amplitude. 
Its effect is clearly visible, as shown in Fig.~\ref{fig4}.
We note that the noise floor in Fig.~\ref{fig4} is at the equivalent of $100\,fN\,mHz^{-1/2}$. 
As a comparison, the typical force of a biological motor molecule is approximately $6.5\,pN$. 
Similarly, the gravitational force exerted by a $75\,kg$ mass on a $1\,g$ test mass at a distance of $1\,m$ is approximately $5\,pN$. 
In both cases, the forces are approximately $50$ times larger than the sensitivity shown here. 
We note that the small force measured here is purely due to the radiation pressure of light. 
While in some cases bolometric effects (e.g., uneven heating and thermal expansion of the device) can also cause a similar deflection, here the effect is negligible. 
This is verified by calibrating the light force on the system with electrostatic forces, by adjusting the voltages applied on the electrodes. 
\begin{figure}                                          
\includegraphics{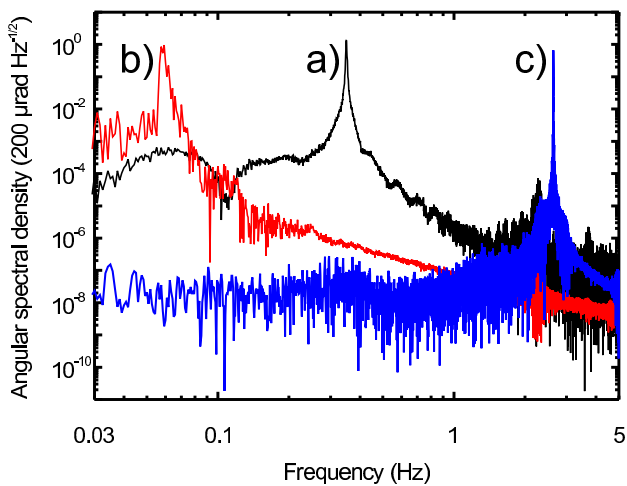}
\caption{Dynamic control of the oscillator's natural frequency. (a) Uncontrolled oscillator, $0.36\,Hz$, (b) positive $P$, $0.06\,Hz$, and (c) negative $P$, $\sim2.6\,Hz$. The oscillator is excited externally for measurement.}
\label{fig5}
\end{figure}
The same levels of rotation of the torsion balance were observed when the same torque was applied, whether by light or by electrostatic force.
  
Finally, we demonstrate the dynamic control of the oscillator's natural frequency. 
In Eq.~\ref{equ1}, if we add an \lq\lq elastic\rq\rq~acceleration: $\alpha_{P}(t)=P\dot\theta$, where $P$ is a constant, the effective natural frequency will become $\omega_{0}=\sqrt{(\tau-P)/I}$. 
Depending on the choice of the sign and magnitude of the control parameter $P$, a wide range of effective natural frequencies can be obtained. 
In Fig.~\ref{fig5}, we show the result of this experiment.
We note that the seismic noise spectra vary from site to site.
With the method demonstrated here, it becomes possible to tailor the oscillator's frequency response and resonance to seek a \lq quiet\rq~frequency band where external noise is minimal, similar to that used in the VIRGO project~\cite{ballardin2001}.
A similar concept was also applied to confine or \lq trap\rq~a mirror aiming at the detection of gravitational waves~\cite{corbitt2_2007}, where the \lq\lq optical spring\rq\rq~force~\cite{sheard2004} is used to raise the natural frequency by over a decade.

In summary, we have demonstrated the dynamic cooling and control of a macroscopic oscillator, using an external, noise-free feedback force. 
Although we used electrostatic forces in the experiment, other means, such as light pressure or magneto-static force can also be applied to control the system. 
We have shown a reduction of the system's equivalent temperature from $\sim10^{5}\,K$, due to the micro-seismic band, to $<300\,mK$. 
To our knowledge, this is the first time that such a great reduction of noise temperature, almost six orders of magnitude, has been achieved. 
The theoretical model we have given, albeit simple, describes the system's behavior very well and affords clear physical insights. 
Using the system, we further showed that very weak forces can be measured with $fN$ sensitivity. 
Furthermore, we demonstrated the control of the oscillator's natural frequency, over a wide range. 
Finally, we note that the dynamic cooling and control method we have developed and reported here is a versatile one. 
It can be applied to experimental studies of other areas such as opto-mechanical multi-stability and parametric processes~\cite{dorsel1983,rugar1991,rokhsari2006,corbitt1_2006} and the \lq optical spring\rq~effects~\cite{corbitt2_2007}. 
We will also examine the applicability of the method in precision measurement of the Newtonian constant ($G$)~\cite{luther1982,quinn2001,schwarz1998} and other applications~\cite{jones1978,smith1999,lamoreaux1997}.

\begin{acknowledgments}
We thank S. Malzer and B. Menegozzi for technical help. We thank Z. H. Lu for helpful discussions.
\end{acknowledgments}


\begin{thebibliography}{32}
\expandafter\ifx\csname natexlab\endcsname\relax\def\natexlab#1{#1}\fi
\expandafter\ifx\csname bibnamefont\endcsname\relax
  \def\bibnamefont#1{#1}\fi
\expandafter\ifx\csname bibfnamefont\endcsname\relax
  \def\bibfnamefont#1{#1}\fi
\expandafter\ifx\csname citenamefont\endcsname\relax
  \def\citenamefont#1{#1}\fi
\expandafter\ifx\csname url\endcsname\relax
  \def\url#1{\texttt{#1}}\fi
\expandafter\ifx\csname urlprefix\endcsname\relax\def\urlprefix{URL }\fi
\providecommand{\bibinfo}[2]{#2}
\providecommand{\eprint}[2][]{\url{#2}}

\bibitem[{\citenamefont{Braginsky and Manukin}(1977)}]{brag1_1977}
\bibinfo{author}{\bibfnamefont{V.~B.} \bibnamefont{Braginsky}}
  \bibnamefont{and} \bibinfo{author}{\bibfnamefont{A.~B.}
  \bibnamefont{Manukin}}, \emph{\bibinfo{title}{Measurement of weak forces in
  physics experiments}} (\bibinfo{publisher}{University of Chicago Press},
  \bibinfo{year}{1977}).

\bibitem[{\citenamefont{Chen and Cook}(1993)}]{chen1_1993}
\bibinfo{author}{\bibfnamefont{Y.~T.} \bibnamefont{Chen}} \bibnamefont{and}
  \bibinfo{author}{\bibfnamefont{A.}~\bibnamefont{Cook}},
  \emph{\bibinfo{title}{Gravitational experiments in the laboratory}}
  (\bibinfo{publisher}{Cambridge University Press}, \bibinfo{year}{1993}).

\bibitem[{\citenamefont{Roll et~al.}(1964)\citenamefont{Roll, Krotkov, and
  Dicke}}]{roll1964}
\bibinfo{author}{\bibfnamefont{P.~G.} \bibnamefont{Roll}},
  \bibinfo{author}{\bibfnamefont{R.}~\bibnamefont{Krotkov}}, \bibnamefont{and}
  \bibinfo{author}{\bibfnamefont{R.~H.} \bibnamefont{Dicke}},
  \bibinfo{journal}{Ann. Phys. (N.Y.)} \textbf{\bibinfo{volume}{26}},
  \bibinfo{pages}{442} (\bibinfo{year}{1964}).

\bibitem[{\citenamefont{Braginsky and Panov}(1972)}]{brag2_1972}
\bibinfo{author}{\bibfnamefont{V.~B.} \bibnamefont{Braginsky}}
  \bibnamefont{and} \bibinfo{author}{\bibfnamefont{V.~I.} \bibnamefont{Panov}},
  \bibinfo{journal}{Sov. Phys. JETP} \textbf{\bibinfo{volume}{34}},
  \bibinfo{pages}{463} (\bibinfo{year}{1972}).

\bibitem[{\citenamefont{Baessler et~al.}(1999)}]{baessler1999}
\bibinfo{author}{\bibfnamefont{S.}~\bibnamefont{Baessler}}
  \bibnamefont{et~al.}, \bibinfo{journal}{Phys. Rev. Lett.}
  \textbf{\bibinfo{volume}{83}}, \bibinfo{pages}{3585} (\bibinfo{year}{1999}).

\bibitem[{\citenamefont{Abramovici et~al.}(1992)}]{abramo1992}
\bibinfo{author}{\bibfnamefont{A.}~\bibnamefont{Abramovici}}
  \bibnamefont{et~al.}, \bibinfo{journal}{Science}
  \textbf{\bibinfo{volume}{256}}, \bibinfo{pages}{325} (\bibinfo{year}{1992}),
  \bibinfo{note}{see also: http://www.ligo.caltech.edu/}.

\bibitem[{\citenamefont{Gillies and Ritter}(1992)}]{gillies1992}
\bibinfo{author}{\bibfnamefont{G.~T.} \bibnamefont{Gillies}} \bibnamefont{and}
  \bibinfo{author}{\bibfnamefont{R.~C.} \bibnamefont{Ritter}},
  \bibinfo{journal}{Rev. Sci. Instrum.} \textbf{\bibinfo{volume}{64}},
  \bibinfo{pages}{283} (\bibinfo{year}{1992}).

\bibitem[{\citenamefont{Chen and Cook}(1990)}]{chen2_1990}
\bibinfo{author}{\bibfnamefont{Y.~T.} \bibnamefont{Chen}} \bibnamefont{and}
  \bibinfo{author}{\bibfnamefont{A.}~\bibnamefont{Cook}},
  \bibinfo{journal}{Class. Quantum Grav.} \textbf{\bibinfo{volume}{7}},
  \bibinfo{pages}{1225} (\bibinfo{year}{1990}).

\bibitem[{\citenamefont{Saulson}(1990)}]{saulson_1990}
\bibinfo{author}{\bibfnamefont{P.~R.} \bibnamefont{Saulson}},
  \bibinfo{journal}{Phys. Rev. D} \textbf{\bibinfo{volume}{42}},
  \bibinfo{pages}{2437} (\bibinfo{year}{1990}).

\bibitem[{\citenamefont{Ritter et~al.}(1999)\citenamefont{Ritter, Winkler, and
  Gillies}}]{ritter1_1999}
\bibinfo{author}{\bibfnamefont{R.~C.} \bibnamefont{Ritter}},
  \bibinfo{author}{\bibfnamefont{L.~I.} \bibnamefont{Winkler}},
  \bibnamefont{and} \bibinfo{author}{\bibfnamefont{G.~T.}
  \bibnamefont{Gillies}}, \bibinfo{journal}{Meas. Sci. Technol.}
  \textbf{\bibinfo{volume}{10}}, \bibinfo{pages}{499} (\bibinfo{year}{1999}).

\bibitem[{\citenamefont{Ritter and Gillies}(1985)}]{ritter2_1985}
\bibinfo{author}{\bibfnamefont{R.~C.} \bibnamefont{Ritter}} \bibnamefont{and}
  \bibinfo{author}{\bibfnamefont{G.~T.} \bibnamefont{Gillies}},
  \bibinfo{journal}{Phys. Rev. A} \textbf{\bibinfo{volume}{31}},
  \bibinfo{pages}{995} (\bibinfo{year}{1985}).

\bibitem[{\citenamefont{Luther and Towler}(1982)}]{luther1982}
\bibinfo{author}{\bibfnamefont{G.~G.} \bibnamefont{Luther}} \bibnamefont{and}
  \bibinfo{author}{\bibfnamefont{W.~R.} \bibnamefont{Towler}},
  \bibinfo{journal}{Phys. Rev. Lett.} \textbf{\bibinfo{volume}{48}},
  \bibinfo{pages}{121} (\bibinfo{year}{1982}).

\bibitem[{\citenamefont{Quinn et~al.}(2001)\citenamefont{Quinn, Speake,
  Richman, Davis, and Picard}}]{quinn2001}
\bibinfo{author}{\bibfnamefont{T.~J.} \bibnamefont{Quinn}},
  \bibinfo{author}{\bibfnamefont{C.~C.} \bibnamefont{Speake}},
  \bibinfo{author}{\bibfnamefont{S.~J.} \bibnamefont{Richman}},
  \bibinfo{author}{\bibfnamefont{R.~S.} \bibnamefont{Davis}}, \bibnamefont{and}
  \bibinfo{author}{\bibfnamefont{A.}~\bibnamefont{Picard}},
  \bibinfo{journal}{Phys. Rev. Lett.} \textbf{\bibinfo{volume}{87}},
  \bibinfo{pages}{111101} (\bibinfo{year}{2001}).

\bibitem[{\citenamefont{Schwarz et~al.}(1998)\citenamefont{Schwarz, Robertson,
  Niebauer, and Faller}}]{schwarz1998}
\bibinfo{author}{\bibfnamefont{J.~P.} \bibnamefont{Schwarz}},
  \bibinfo{author}{\bibfnamefont{D.~S.} \bibnamefont{Robertson}},
  \bibinfo{author}{\bibfnamefont{T.~M.} \bibnamefont{Niebauer}},
  \bibnamefont{and} \bibinfo{author}{\bibfnamefont{J.~E.}
  \bibnamefont{Faller}}, \bibinfo{journal}{Science}
  \textbf{\bibinfo{volume}{282}}, \bibinfo{pages}{2230} (\bibinfo{year}{1998}).

\bibitem[{\citenamefont{Niebauer et~al.}(1995)\citenamefont{Niebauer, Sasagawa,
  Faller, Hilt, and Klopping}}]{niebauer1995}
\bibinfo{author}{\bibfnamefont{T.~M.} \bibnamefont{Niebauer}},
  \bibinfo{author}{\bibfnamefont{G.~S.} \bibnamefont{Sasagawa}},
  \bibinfo{author}{\bibfnamefont{J.~E.} \bibnamefont{Faller}},
  \bibinfo{author}{\bibfnamefont{R.}~\bibnamefont{Hilt}}, \bibnamefont{and}
  \bibinfo{author}{\bibfnamefont{F.}~\bibnamefont{Klopping}},
  \bibinfo{journal}{Metrologia} \textbf{\bibinfo{volume}{32}},
  \bibinfo{pages}{159} (\bibinfo{year}{1995}).

\bibitem[{\citenamefont{Cohadon et~al.}(1999)\citenamefont{Cohadon, Heidmann,
  and Pinard}}]{cohadon1999}
\bibinfo{author}{\bibfnamefont{P.~F.} \bibnamefont{Cohadon}},
  \bibinfo{author}{\bibfnamefont{A.}~\bibnamefont{Heidmann}}, \bibnamefont{and}
  \bibinfo{author}{\bibfnamefont{M.}~\bibnamefont{Pinard}},
  \bibinfo{journal}{Phys. Rev. Lett.} \textbf{\bibinfo{volume}{83}},
  \bibinfo{pages}{3174} (\bibinfo{year}{1999}).

\bibitem[{\citenamefont{Metzger and Karrai}(2004)}]{metzger2004}
\bibinfo{author}{\bibfnamefont{C.~H.} \bibnamefont{Metzger}} \bibnamefont{and}
  \bibinfo{author}{\bibfnamefont{K.}~\bibnamefont{Karrai}},
  \bibinfo{journal}{Nature} \textbf{\bibinfo{volume}{432}},
  \bibinfo{pages}{1002} (\bibinfo{year}{2004}).

\bibitem[{\citenamefont{Gigan et~al.}(2006)}]{gigan2006}
\bibinfo{author}{\bibfnamefont{S.}~\bibnamefont{Gigan}} \bibnamefont{et~al.},
  \bibinfo{journal}{Nature} \textbf{\bibinfo{volume}{444}}, \bibinfo{pages}{67}
  (\bibinfo{year}{2006}).

\bibitem[{\citenamefont{Arcizet
  et~al.}(2006{\natexlab{a}})\citenamefont{Arcizet, Cohadon, Briant, Pinard,
  and Heidmann}}]{arcizet1_2006}
\bibinfo{author}{\bibfnamefont{O.}~\bibnamefont{Arcizet}},
  \bibinfo{author}{\bibfnamefont{P.~F.} \bibnamefont{Cohadon}},
  \bibinfo{author}{\bibfnamefont{T.}~\bibnamefont{Briant}},
  \bibinfo{author}{\bibfnamefont{M.}~\bibnamefont{Pinard}}, \bibnamefont{and}
  \bibinfo{author}{\bibfnamefont{A.}~\bibnamefont{Heidmann}},
  \bibinfo{journal}{Nature} \textbf{\bibinfo{volume}{444}}, \bibinfo{pages}{71}
  (\bibinfo{year}{2006}{\natexlab{a}}).

\bibitem[{\citenamefont{Kleckner and Bouwmeester}(2006)}]{kleckner2006}
\bibinfo{author}{\bibfnamefont{D.}~\bibnamefont{Kleckner}} \bibnamefont{and}
  \bibinfo{author}{\bibfnamefont{D.}~\bibnamefont{Bouwmeester}},
  \bibinfo{journal}{Nature} \textbf{\bibinfo{volume}{444}}, \bibinfo{pages}{75}
  (\bibinfo{year}{2006}).

\bibitem[{\citenamefont{Arcizet et~al.}(2006{\natexlab{b}})}]{arcizet2_2006}
\bibinfo{author}{\bibfnamefont{O.}~\bibnamefont{Arcizet}} \bibnamefont{et~al.},
  \bibinfo{journal}{Phys. Rev. Lett.} \textbf{\bibinfo{volume}{97}},
  \bibinfo{pages}{133601} (\bibinfo{year}{2006}{\natexlab{b}}).

\bibitem[{\citenamefont{Lorrain}(1991)}]{lorrain1991}
\bibinfo{author}{\bibfnamefont{P.}~\bibnamefont{Lorrain}},
  \bibinfo{journal}{Opt. Laser Eng.} \textbf{\bibinfo{volume}{15}},
  \bibinfo{pages}{197} (\bibinfo{year}{1991}).

\bibitem[{\citenamefont{Ballardin et~al.}(2001)}]{ballardin2001}
\bibinfo{author}{\bibfnamefont{G.}~\bibnamefont{Ballardin}}
  \bibnamefont{et~al.}, \bibinfo{journal}{Rev. Sci. Instrum.}
  \textbf{\bibinfo{volume}{72}}, \bibinfo{pages}{3643} (\bibinfo{year}{2001}).

\bibitem[{\citenamefont{Corbitt et~al.}(2007)}]{corbitt2_2007}
\bibinfo{author}{\bibfnamefont{T.}~\bibnamefont{Corbitt}} \bibnamefont{et~al.},
  \bibinfo{journal}{Phys. Rev. Lett.} \textbf{\bibinfo{volume}{98}},
  \bibinfo{pages}{150802} (\bibinfo{year}{2007}).

\bibitem[{\citenamefont{Sheard et~al.}(2004)\citenamefont{Sheard, Gray,
  Mow-Lowry, McClelland, and Whitcomb}}]{sheard2004}
\bibinfo{author}{\bibfnamefont{B.~S.} \bibnamefont{Sheard}},
  \bibinfo{author}{\bibfnamefont{M.~B.} \bibnamefont{Gray}},
  \bibinfo{author}{\bibfnamefont{C.~M.} \bibnamefont{Mow-Lowry}},
  \bibinfo{author}{\bibfnamefont{D.~E.} \bibnamefont{McClelland}},
  \bibnamefont{and} \bibinfo{author}{\bibfnamefont{S.~E.}
  \bibnamefont{Whitcomb}}, \bibinfo{journal}{Phys. Rev. A}
  \textbf{\bibinfo{volume}{69}}, \bibinfo{pages}{051801}
  (\bibinfo{year}{2004}).

\bibitem[{\citenamefont{Dorsel et~al.}(1983)\citenamefont{Dorsel, McCullen,
  Meystre, Vignes, and Walther}}]{dorsel1983}
\bibinfo{author}{\bibfnamefont{A.}~\bibnamefont{Dorsel}},
  \bibinfo{author}{\bibfnamefont{J.~D.} \bibnamefont{McCullen}},
  \bibinfo{author}{\bibfnamefont{P.}~\bibnamefont{Meystre}},
  \bibinfo{author}{\bibfnamefont{E.}~\bibnamefont{Vignes}}, \bibnamefont{and}
  \bibinfo{author}{\bibfnamefont{H.}~\bibnamefont{Walther}},
  \bibinfo{journal}{Phys. Rev. Lett.} \textbf{\bibinfo{volume}{51}},
  \bibinfo{pages}{1550} (\bibinfo{year}{1983}).

\bibitem[{\citenamefont{Rugar and Gruetter}(1991)}]{rugar1991}
\bibinfo{author}{\bibfnamefont{D.}~\bibnamefont{Rugar}} \bibnamefont{and}
  \bibinfo{author}{\bibfnamefont{P.}~\bibnamefont{Gruetter}},
  \bibinfo{journal}{Phys. Rev. Lett.} \textbf{\bibinfo{volume}{67}},
  \bibinfo{pages}{699} (\bibinfo{year}{1991}).

\bibitem[{\citenamefont{Rokhsari et~al.}(2006)\citenamefont{Rokhsari,
  Kippenberg, Carmon, and Vahala}}]{rokhsari2006}
\bibinfo{author}{\bibfnamefont{H.}~\bibnamefont{Rokhsari}},
  \bibinfo{author}{\bibfnamefont{T.~J.} \bibnamefont{Kippenberg}},
  \bibinfo{author}{\bibfnamefont{T.}~\bibnamefont{Carmon}}, \bibnamefont{and}
  \bibinfo{author}{\bibfnamefont{K.~J.} \bibnamefont{Vahala}},
  \bibinfo{journal}{IEEE J. Quant. Elec.} \textbf{\bibinfo{volume}{12}},
  \bibinfo{pages}{96} (\bibinfo{year}{2006}).

\bibitem[{\citenamefont{Corbitt et~al.}(2006)\citenamefont{Corbitt, Ottaway,
  Innerhofer, Pelc, and Mavalvala}}]{corbitt1_2006}
\bibinfo{author}{\bibfnamefont{T.}~\bibnamefont{Corbitt}},
  \bibinfo{author}{\bibfnamefont{D.}~\bibnamefont{Ottaway}},
  \bibinfo{author}{\bibfnamefont{E.}~\bibnamefont{Innerhofer}},
  \bibinfo{author}{\bibfnamefont{J.}~\bibnamefont{Pelc}}, \bibnamefont{and}
  \bibinfo{author}{\bibfnamefont{N.}~\bibnamefont{Mavalvala}},
  \bibinfo{journal}{Phys. Rev. A} \textbf{\bibinfo{volume}{74}},
  \bibinfo{pages}{021802} (\bibinfo{year}{2006}).

\bibitem[{\citenamefont{Jones and Leslie}(1978)}]{jones1978}
\bibinfo{author}{\bibfnamefont{R.~V.} \bibnamefont{Jones}} \bibnamefont{and}
  \bibinfo{author}{\bibfnamefont{B.}~\bibnamefont{Leslie}},
  \bibinfo{journal}{Proc. R. Soc. Lond. A} \textbf{\bibinfo{volume}{360}},
  \bibinfo{pages}{347} (\bibinfo{year}{1978}).

\bibitem[{\citenamefont{Smith et~al.}(1999)}]{smith1999}
\bibinfo{author}{\bibfnamefont{G.~L.} \bibnamefont{Smith}}
  \bibnamefont{et~al.}, \bibinfo{journal}{Phys. Rev. D}
  \textbf{\bibinfo{volume}{61}}, \bibinfo{pages}{022001}
  (\bibinfo{year}{1999}).

\bibitem[{\citenamefont{Lamoreaux}(1997)}]{lamoreaux1997}
\bibinfo{author}{\bibfnamefont{S.~K.} \bibnamefont{Lamoreaux}},
  \bibinfo{journal}{Phys. Rev. Lett.} \textbf{\bibinfo{volume}{78}},
  \bibinfo{pages}{5} (\bibinfo{year}{1997}).

\end{thebibliography}
\end{document}